\documentclass[11pt]{article}

\usepackage{bm}
\usepackage[normalem]{ulem}
\usepackage[utf8]{inputenc}
\usepackage[T1]{fontenc}
\usepackage{mathptmx}
\usepackage{etoolbox}
\usepackage{xcolor}
\usepackage{physics}

\usepackage[a4paper, margin=0.9in]{geometry}
\usepackage{authblk}
\usepackage{graphicx}
\usepackage{dcolumn}
\usepackage{amsmath}
\usepackage{textcomp}
\usepackage{verbatim}
\usepackage{float}
\usepackage[symbol]{footmisc}
\setfnsymbol{wiley}
\usepackage{hyperref}
\hypersetup{
    colorlinks=true,
    linkcolor=blue,
    filecolor=blue,
    urlcolor=blue,
    citecolor = blue,
    pdfauthor=author,
}

\usepackage{multirow}
\usepackage[T1]{fontenc}
\usepackage[normalem]{ulem}
\usepackage[numbers,sort&compress]{natbib}

\begin{document}

\title{Polarization-controlled Brillouin scattering in elliptical optophononic resonators}

\author[1]{A. Rodriguez \thanks{These authors contributed equally.}}
\author[*1]{E. Mehdi}
\author[1]{Priya}
\author[1]{E. R. Cardozo de Oliveira}
\author[1]{M. Esmann}
\author[1]{N. D. Lanzillotti-Kimura \footnote{daniel.kimura@c2n.upsaclay.fr}}

\affil[1]{Université Paris-Saclay, CNRS, Centre de Nanosciences et de Nanotechnologies, 91120 Palaiseau, France}

\date{}
\maketitle
\begin{abstract}
The fast-growing development of optomechanical applications has motivated advancements in Brillouin scattering research. In particular, the study of high frequency acoustic phonons at the nanoscale is interesting due to large range of interactions with other excitations in matter. However, standard Brillouin spectroscopy schemes rely on fixed wavelength filtering, which limits the usefulness for the study of tunable optophononic resonators.
It has been recently demonstrated that elliptical optophononic micropillar resonators induce different energy-dependent polarization states for the Brillouin and the elastic Rayleigh scattering, and that a polarization filtering setup could be implemented to increase the contrast between the inelastic and elastic scattering of the light. 
An optimal filtering configuration can be reached when the polarization states of the laser and the Brillouin signal are orthogonal from each other. 
In this work, we theoretically investigate the parameters of such polarization-based filtering technique to enhance the efficiency of Brillouin scattering detection. 
For the filtering optimization, we explore the initial wavelength and polarization state of the incident laser, as well as in the ellipticity of the micropillars, and reach an almost optimal configuration for nearly background-free Brillouin detection. 
Our findings are one step forward on the efficient detection of Brillouin scattering in nanostructures for potential applications in fields such as optomechanics and quantum communication.
\end{abstract}

\maketitle 

\section{Introduction}

The engineering of nanomechanical systems has major impact in heat transport management \cite{Florez2022,Braun2022}, opto-acoustic cooling \cite{blazquez2024}, quantum correlations mediated by single phonons \cite{anderson_two-color_2018}, and data storage \cite{stillerCoherentlyRefreshingHypersonic2020a,merklein_-chip_2020} among others.  
These advances motivate the development of new techniques to investigate high frequency phonons in the GHz range with high precision \cite{volzNanophononicsStateArt2016,Priya2023,Cardozo2023, Ortiz2020,machadoGenerationPropagationSuperhighFrequency2019a}. 
For instance, Brillouin scattering, the inelastic scattering of light with acoustic phonons, has experienced a renewed interest in the communities of nanophononics and opto-nano-mechanics \cite{Eggleton2019,kargarAdvancesBrillouinMandelstam2021a}.

The intrinsic properties of the materials govern the Brillouin scattering selection rules and constrain the energy, direction, and polarization of the Brillouin signal \cite{Yu_Cardona_2010}. 
In addition, the overlap between optical and acoustic fields determines the efficiency of the inelastic scattering process. 
In this context, GaAs/AlAs optophononic systems allow to simultaneously control  the confinement of ultrahigh-frequency acoustic phonons and photons in the near-infrared \cite{Ortiz2021,Fainstein2013,anguianoMicropillarResonatorsOptomechanics2017}. 
The overlap of the optical and acoustical fields in such optophononic cavity leads to the enhancement of the Brillouin scattering signals \cite{lanzillotti-kimura_resonant_2009,rodriguez_fiber-based_2021}.
Three-dimensional confinement of light and sound can be achieved in semiconductor multilayered devices etched as micropillars \cite{esmann_brillouin_2019,lamberti_optomechanical_2017}. Such cavity systems receive attention as they can be, for example, integrated with quantum dot to develop single photon sources \cite{Somaschi2016,Wang2019,Bakker_2015}. 
To detect the spontaneous Brillouin scattering signals, a particularly stringent filtering of the elastically scattered laser is necessary.  
Therefore, new techniques of tunable Brillouin spectroscopy are developed to allow the efficient filtering of the scattered signal \cite{Rozas2014, rodriguez_fiber-based_2021,esmann_brillouin_2019}. 

Here, we study elliptical micropillars, that enable the control of the Brillouin scattering polarization selection rules \cite{rodriguez_brillouin_2023}. 
Indeed, the anisotropy of the micropillar cross-section lifts the degeneracy of the optical cavity modes, in energy and polarization \cite{Rakher2009,Gayral1998,Whittaker2017}. 
A wavelength-dependent polarization rotation is then induced by the birefringence of the elliptical microcavity \cite{Hilaire2018, Androvi2019}. 
Contrary to standard polarization selection rules in GaAs grown along the [001] crystalline direction, the Brillouin scattering polarization in elliptical micropillars can be rotated differently than the polarization of the excitation laser reflected by the optical cavity.

In a previous work \cite{rodriguez_brillouin_2023}, we reported the theoretical and experimental demonstration of the manipulation of the Brillouin scattering polarization with an excitation laser polarized along the linear diagonal polarization. In this work, we theoretically introduce different parameters to control the polarization of the Brillouin scattering.We demonstrate the possibility to efficiently filter out the reflected laser by reaching an almost cross-polarization configuration between the laser and the Brillouin signal.

\section{Elliptical optophononic micropillar cavities}

\begin{figure*}[h!]
\centering
\includegraphics[scale=0.35]{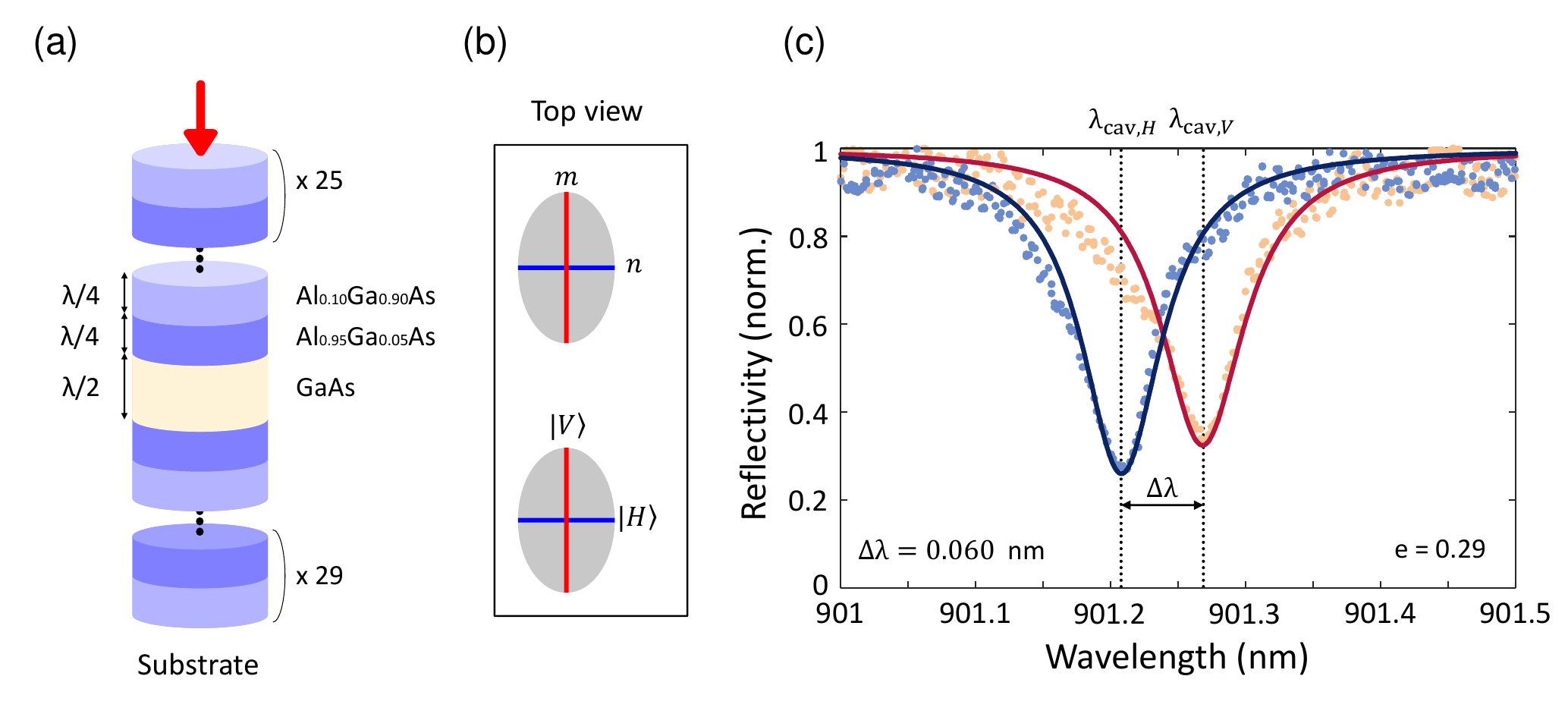}
\caption{(a) Schematic of the micropillar vertical layer structure with two distributed Bragg reflectors (blue) surrounding a resonant spacer (light yellow).    
(b) Schematic of the micropillar cross-section with $m$ and $n$ the major and minor axes of the elliptical structure. The two linear polarizations $H$ and $V$ are the two fundamental optical eigenmodes of the micropillar. 
(c) Measured (dots) and fitted (solid line) optical reflectivity of an elliptical pillar with an ellipticity of e = 0.29.
The blue (red) line corresponds to $\ket{H}$ ($\ket{V}$).}
\label{figure1}
\end{figure*}

We theoretically study here an optophononic microcavity based on distributed Bragg reflectors (DBRs), grown on a (001)-oriented GaAs substrate by molecular-beam epitaxy.
We use experimental data from the same sample as in \cite{rodriguez_brillouin_2023} to base our study on experimentally realistic conditions.
A resonant spacer of optical path length of $\lambda$/2 at a resonance wavelength of $\lambda \sim 900\,$nm is enclosed by the two DBRs.
The top (bottom) optical DBR is formed by 25 (29) periods of $\mathrm{Ga}_\mathrm{0.9}\mathrm{Al}_\mathrm{0.1}\mathrm{As}$/$\mathrm{Ga}_\mathrm{0.05}\mathrm{Al}_\mathrm{0.95}\mathrm{As}$ bilayers. 
Elliptical micropillars as schematised in Fig. \ref{figure1}a are fabricated from this planar optophononic cavity by optical lithography followed by Inductively Coupled Plasma Etching (ICP). 
A schematic of the elliptical cross-section is shown in Fig. \ref{figure1}b where the major and minor axes are indicated (\textit{m} and \textit{n}, respectively).
The optical cavity presents two eigenmodes with orthogonal linear polarization and two nondegenerate eigenenergies, $\omega_{\mathrm{cav}}^H$ and $\omega_{\mathrm{cav}}^V$ (considering $\hbar =1$), due to the anisotropy of the micropillar cross-section.
We define as well, for each mode, its corresponding wavelength $\lambda_{\mathrm{cav}}^{H/V}$. 
These two optical modes correspond to the horizontal ($H$) and vertical ($V$) polarizations.

When considering a linearly polarized incident laser beam, $\ket{\psi_{\mathrm{in}}} $,
the output polarization $\ket{\psi_{\mathrm{out}}}$ of the laser reflected by the cavity is rotated as a function of the laser wavelength $\lambda_{\mathrm{laser}}$ due to the cavity birefringence \cite{Hilaire2018, Sun2016}. 
Both polarization states can be defined as follows:
\begin{equation}
\begin{split}
    & \ket{\psi_{\mathrm{in}}} = \dfrac{1}{\sqrt{|b_{\mathrm{in}}^H|^2 + |b_{\mathrm{in}}^V|^2}}(b_{\mathrm{in}}^H\ket{H} + b_{\mathrm{in}}^V\ket{V}), \\ 
    & \ket{\psi_{\mathrm{out}}} = \dfrac{1}{\sqrt{|b_{\mathrm{out}}^H|^2 + |b_{\mathrm{out}}^V|^2}}(b_{\mathrm{out}}^H\ket{H} + b_{\mathrm{out}}^V\ket{V}),
\end{split}  
\end{equation}
where $b_{\mathrm{in}}^{H/V}$ and $b_{\mathrm{out}}^{H/V}$ are the average values of the polarization dependent input and output external field operators \cite{rodriguez_brillouin_2023}. 
These operators are defined with the input-output relation \cite{walls_quantum_2008}:
\begin{equation} \label{Eq_input_output}
    b_{\mathrm{out}}^{H/V} = b_{\mathrm{in}}^{H/V} + \sqrt{\kappa_{\mathrm{top},H/V}} a_{H/V}
\end{equation}
with $\kappa_{\mathrm{top},H/V}$ being the top DBR damping rate and $a_{H/V}$ the average value of the intracavity field operator. 
The reflected light, represented by $b_{\mathrm{out}}^{H/V}$, is the result of the interference between the light emerging from the cavity and the light directly reflected by the top DBR. 
For a given input polarization $\ket{\psi_{\mathrm{in}}}$, the response in polarization of the cavity can be calculated with the polarization-dependent complex reflection coefficients $r_H$ and $r_V$ \cite{Mehdi2024}:
\begin{equation}
\begin{split}
    r_{H/V} = &  \dfrac{b_{\mathrm{out}}^{H/V}}{b_{\mathrm{in}}^{H/V}} \\
    = & 1 - 2\eta_{\mathrm{top},H/V} \dfrac{1}{1-\dfrac{2i(\omega_{\mathrm{in}}-\omega_{\mathrm{cav}}^{H/V})}{\kappa_{H/V}}},
    \label{equation_reflectivity_coef}
\end{split}
\end{equation}
where $\omega_{\mathrm{in}}$ and $\omega_{\mathrm{cav},H/V}$ are the energies of the incident laser and the cavity modes, respectively. 
The output coupling efficiency through the top DBR $\eta_{\mathrm{top},H/V}$ is defined as $\eta_{\mathrm{top},H/V} = \kappa_{\mathrm{top},H/V} / \kappa_{H/V}$ with $\kappa_{H/V}$ the total cavity damping rate including leakage through the top and bottom DBRs, as well as the sidewall losses.

To theoretically study the polarization rotation induced by the elliptical cavities, we use the experimental  optical cavity parameters of a given micropillar.
A micropillar of ellipticity $e = \sqrt{m/n} - 1 = 0.29$ is chosen from the sample used in reference \cite{rodriguez_brillouin_2023}. 
The two optical cavity modes can be then characterized with reflectivity measurements as a function of the incident laser wavelength as shown in Fig. \ref{figure1}c.
To measure the reflectivity spectra of the micropillar cavity, the polarization of the excitation laser is set to match one of the cavity eigenmodes i.e., $\ket{\psi_{\mathrm{in}}} = \ket{H/V}$.  
The experimental reflectivities (dots) of the $H$ mode (blue) and the $V$ mode (red) are fitted (line) with the optical cavity parameters ($\lambda_{\mathrm{cav},H/V}$, $\eta_{\mathrm{top},H/V}$ and $\kappa_{H/V}$) as the cavity reflectivity is directly deduced from $|r_{H/V}|^2$ for each mode.

\section{Polarization rotation induced by the elliptical cavities}

\begin{figure*}[h!]
\centering
\includegraphics[scale=0.19]{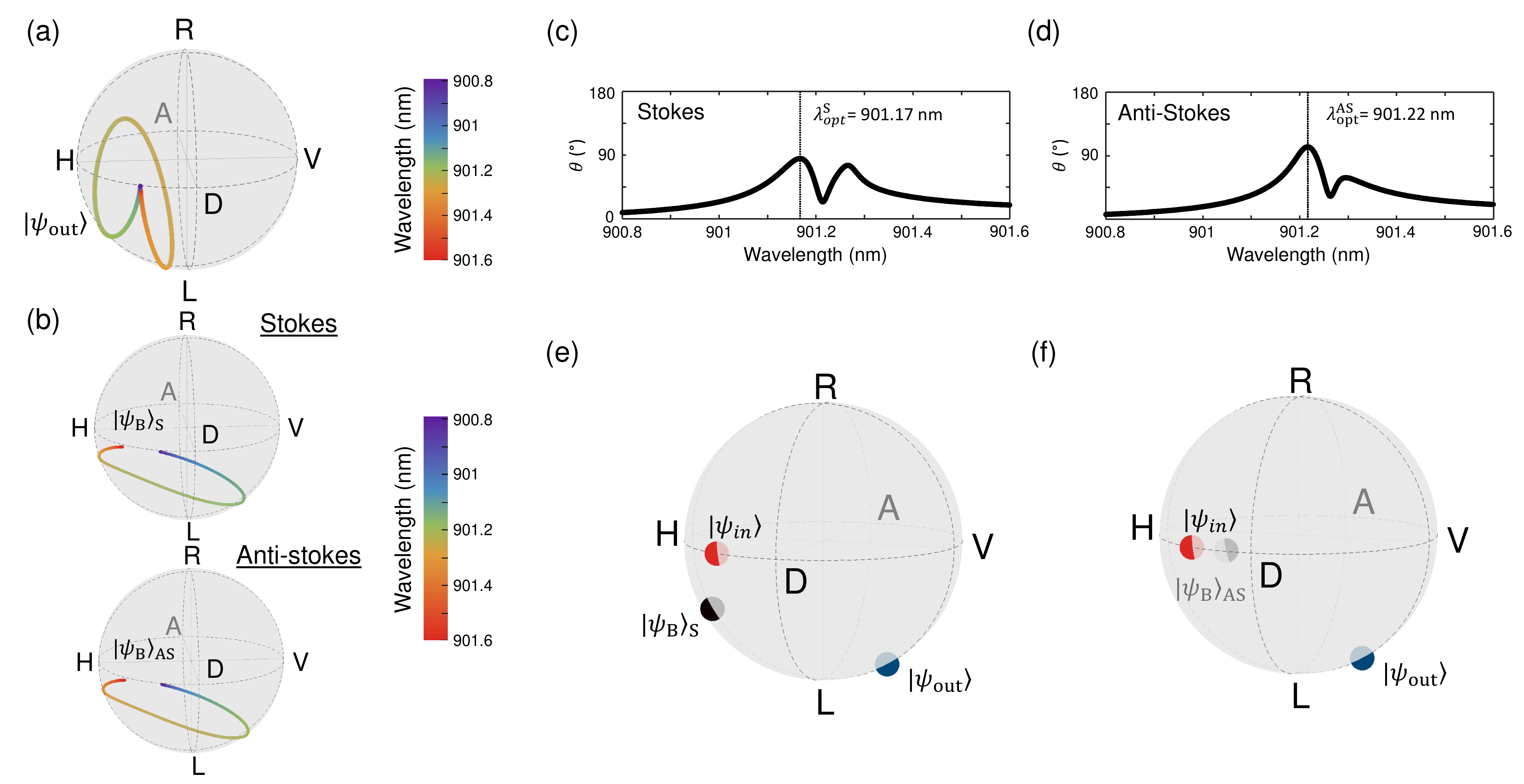}
\caption{(a) Simulated polarization rotation of the polarization state $\ket{\psi_{\mathrm{out}}}$ induced by the cavity birefringence  displayed in the Poincaré sphere as a function of the input laser wavelength indicated by the colorscale. 
(b) Simulated polarization states of the Stokes and anti-Stokes signal of the Brillouin scattering at $18\,$GHz, $\ket{\psi_{\mathrm{B}}}_{\mathrm{S}}$ and $\ket{\psi_{\mathrm{B}}}_{\mathrm{AS}}$, as a function of the input laser wavelength. 
(c) and (d) Angle in the Poincaré sphere between the Stokes (resp. anti-Stokes) signal polarization state and the laser polarization rotated by the cavity as a function of the laser wavelength. 
The optimal wavelength $\lambda_{\mathrm{opt}}^{\mathrm{S}}$ (resp. $\lambda_{\mathrm{opt}}^{\mathrm{AS}}$) corresponds to the maximal angle between $\ket{\psi_{\mathrm{B}}}_{\mathrm{S}}$ (resp. $\ket{\psi_{\mathrm{B}}}_{\mathrm{AS}}$) and $\ket{\psi_{\mathrm{out}}}$.
(e) and (f) Representation in the Poincaré sphere of the polarization states $\ket{\psi_{\mathrm{in}}}$, $\ket{\psi_{\mathrm{B}}}_{\mathrm{S}}$ (resp. $\ket{\psi_{\mathrm{B}}}_{\mathrm{AS}}$) and $\ket{\psi_{\mathrm{out}}}$ at the optimal wavelength $\lambda_{\mathrm{opt}}^{\mathrm{S}}$ (resp. $\lambda_{\mathrm{opt}}^{\mathrm{AS}}$).}
\label{figure2}
\end{figure*}

For an arbitrary linear input polarization state $\ket{\psi_{\mathrm{in}}}$, we calculate the polarization rotation induced by the considered micropillar cavity with Eq. (\ref{equation_reflectivity_coef}) and plot it in the Poincaré sphere as a function of the input laser wavelength (see colorscale) as shown in Fig. \ref{figure2}a.
For each laser wavelength, the Stokes parameters of the output state $\ket{\psi_{\mathrm{out}}}$ are calculated and give the coordinates of the polarization state in the Poincaré sphere. 
Far from the cavity resonance, there is no polarization rotation, i.e., $\ket{\psi_{\mathrm{out}}} = \ket{\psi_{\mathrm{in}}}$, while $\ket{\psi_{\mathrm{out}}}$ forms a closed path on the surface of the Poincaré sphere when in resonance with the cavity, assuming a perfect mode matching between the input beam and optical cavity. 
The trajectory of the output polarization state as a function of the input laser wavelength depends directly on the cavity parameters and the laser energy $\omega_{\mathrm{in}}$.   

The simulations of the output polarization states are based on the Jones matrices formalism \cite{goldstein_polarized_2017, rodriguez_brillouin_2023}.
The cavity parameters deduced from the fit of the cavity reflectivity spectra are used as input for the Jones matrices to calculate the polarization rotation induced by the cavity.

The polarization of the Brillouin scattering field, inside the GaAs cavity spacer, is calculated for every value of the excitation laser wavelength. 
Its polarization amplitude is the same as the intracavity polarization state of the input laser but at a different energy $\omega_{\mathrm{B}}$.
The Brillouin scattering polarization state at the output of the cavity, $\ket{\psi_{\mathrm{B}}}_{\mathrm{B/AS}}$, is thus defined as:
\begin{equation}
    \ket{\psi_{\mathrm{B}}}_{\mathrm{B/AS}} = \dfrac{1}{\sqrt{|b_{\mathrm{B}}^{H}|^2 + |b_{\mathrm{B}}^{V}|^2}}(b_{\mathrm{B}}^{H}\ket{H} + b_{\mathrm{B}}^{V}\ket{V}),
\end{equation}
with $b_{\mathrm{B}}^{H}$ the average value of the Brillouin scattering operator defined as follow:
\begin{equation} \label{Eq_b_Brillouin}
\begin{split}
    &b_{\mathrm{B}}^{H} \propto  \\
    &\dfrac{1}{1-\dfrac{2i(\omega_{\mathrm{B}} - \omega_{\mathrm{cav}}^{H/V})}{\kappa_{H/V}}} \cdot \dfrac{1}{1-\dfrac{2i(\omega_{\mathrm{in}} - \omega_{\mathrm{cav}}^{H/V})}{\kappa_{H/V}}} \cdot b_{\mathrm{in}}^{H/V}
\end{split}
\end{equation}
The first term depends on the frequency of the Brillouin signal $\omega_{\mathrm{B}}$. 
In our case, we consider only the first harmonic at $\pm 18\,$GHz which corresponds to the confined acoustic mode in the spacer. 
The second term depends on the incoming laser frequency $\omega_{\mathrm{in}}$ and its polarization with $b_{\mathrm{in}}^{H/V}$.
The Brillouin scattering polarization state at the output of the cavity is then calculated as a function of the incoming laser wavelength for the Stokes and the anti-Stokes signals at $18\,$GHz and plotted in the Poincaré sphere in the Fig. \ref{figure2}b. 
The behaviour of the Brillouin scattering polarization is different from the one of the light reflected by the cavity. 
The Brillouin signal can therefore be separated from the reflected laser by polarization filtering. 

From the Stokes parameters describing the reflected laser and the Brillouin scattering polarization states, the angle $\theta$ between the polarization states $\ket{\psi_{\mathrm{out}}}$ and $\ket{\psi_{\mathrm{B}}}_{\mathrm{S/AS}}$ in the Poincaré sphere is calculated for the Stokes and anti-Stokes signals as follows:
\begin{equation}
    \theta = 2 \mathrm{arcsin}(\dfrac{\Delta}{2}),
\end{equation}
with $\Delta = \sqrt{| s_{HV}^{\mathrm{B}} - s_{HV}^{\mathrm{out}}|^2 + |s_{DA}^{\mathrm{B}} - s_{DA}^{\mathrm{out}}|^2 + | s_{RL}^{\mathrm{B}} - s_{RL}^{\mathrm{out}}|^2}$.  
The angle $\theta$ is plotted in Fig. \ref{figure2}c and \ref{figure2}d as a function of the wavelength of the input laser for the Stokes and anti-Stokes signals respectively. 
The ideal filtering condition of the reflected laser corresponds to the two polarization states $\ket{\psi_{\mathrm{out}}}$ and $\ket{\psi_{\mathrm{B}}}_{\mathrm{S/AS}}$ orthogonal to each other, i.e., giving two opposite points in the Poincaré sphere.
Under these conditions, with an angle of $\theta = 180^{\circ}$ between the two polarization states in the Poincaré sphere, the reflected laser can be entirely filtered while getting all of the Brillouin signal by 
selecting only the polarization along $\ket{\psi_{\mathrm{B}}}_{\mathrm{S/AS}}$. 
A maximal filtering of the reflected laser can however be achieved when the Brillouin signal polarization state is the furthest from the reflected laser in the Poincaré sphere, corresponding to the maximal angle $\theta_{\mathrm{max}}$.   
In Fig. \ref{figure2}c and \ref{figure2}d, the maximal angles reached for the Stokes and anti-Stokes signals, $\theta_{\mathrm{max}}^{\mathrm{S}} = 85^{\circ}$ and $\theta_{\mathrm{max}}^{\mathrm{AS}} = 102 ^{\circ}$, are indicated with the corresponding wavelength $\lambda_{\mathrm{opt}}^{\mathrm{S}}$ and $\lambda_{\mathrm{opt}}^{\mathrm{AS}}$.
In this situation, the polarization of the Stokes and anti-Stokes signal at $\pm 18\,$GHz are not orthogonal to the reflected laser polarization as the maximal angle does not reach the value of $180^{\circ}$.  
The polarization states $\ket{\psi_{\mathrm{out}}}$ and $\ket{\psi_{\mathrm{B}}}_{\mathrm{S/AS}}$ at the wavelength $\lambda_{\mathrm{opt}}^{\mathrm{S}}$ and $\lambda_{\mathrm{opt}}^{\mathrm{AS}}$ are now compared in the Poincaré sphere in Fig. \ref{figure2}e and \ref{figure2}f, where the input polarization state $\ket{\psi_{\mathrm{in}}}$ is plotted as well. 
It allows the direct comparison of the polarization states $\ket{\psi_{\mathrm{B}}}_{\mathrm{S/AS}}$ and $\ket{\psi_{\mathrm{out}}}$ as they are not opposite to each other in the sphere.
A cross-polarization scheme, adjusted to the reflected laser polarization, would allow to completely eliminated the laser. 
However, the Brillouin signal will be highly attenuated in this situation, as its polarization is not orthogonal to the laser.

For this given input polarization state $\ket{\psi_{\mathrm{in}}}$, the optimal condition reached between the polarization states $\ket{\psi_{\mathrm{B}}}_{\mathrm{S/AS}}$ and $\ket{\psi_{\mathrm{out}}}$ is far from ideal as the two polarization states are not opposite to each other in the Poincaré sphere.
However, these two polarization states depend by definition (see Eq.(\ref{Eq_input_output}) and Eq.(\ref{Eq_b_Brillouin})) on the average, i.e., expectation values, of the polarization dependent external input field operators $b_{\mathrm{in}}^{H/V}$, i.e., they depend on the input polarization state $\ket{\psi_{\mathrm{in}}}$.  
Thus the angle $\theta$ can be furthermore optimised by changing the input polarization state. \\

\section{Optimization of the filtering}

\begin{figure}[h!]
\centering
\includegraphics[scale=0.20]{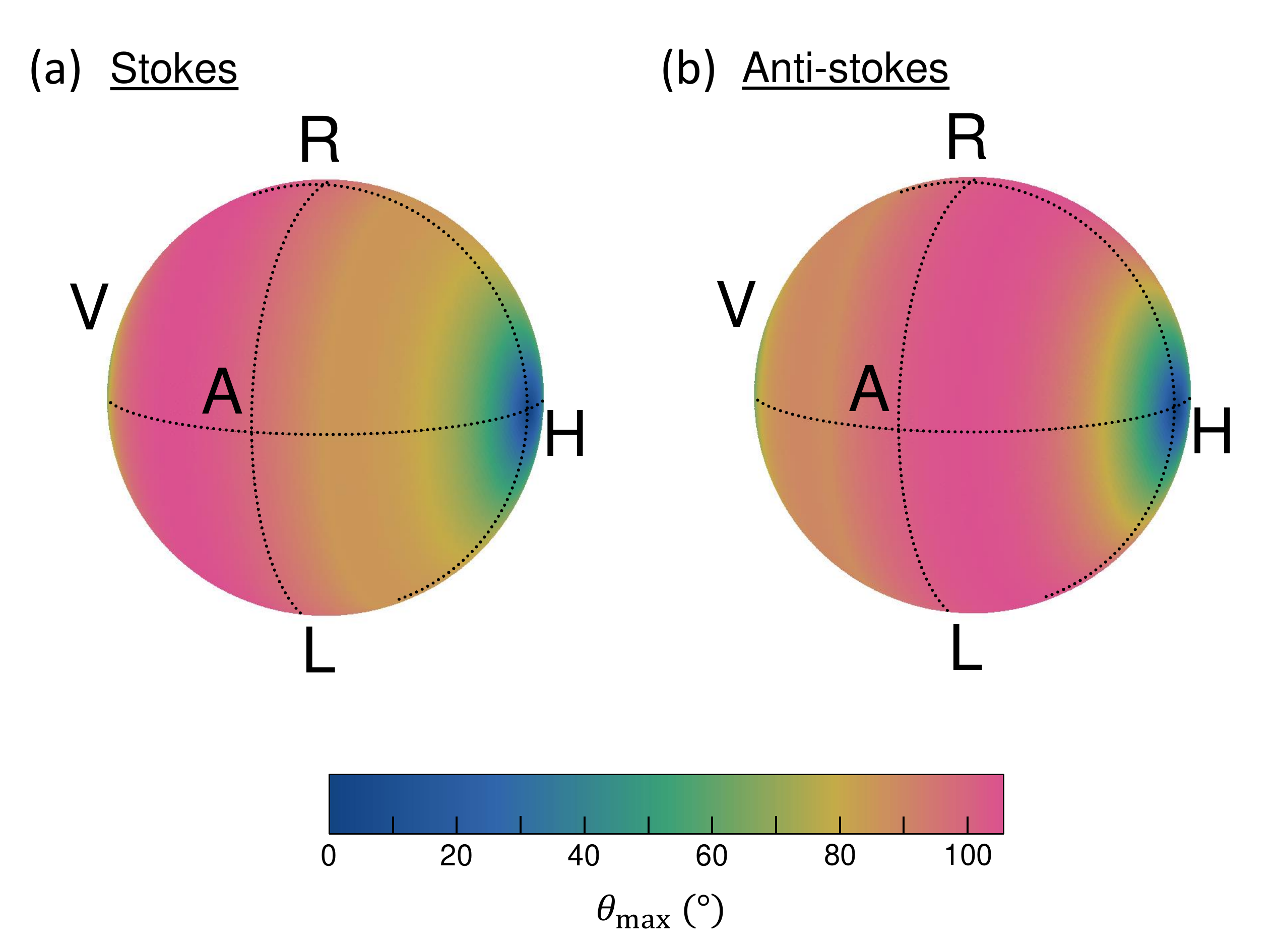}
\caption{Maximal angle $\theta_{\mathrm{max}}$ in the Poincaré sphere between the polarization states of the reflected laser $\ket{\psi_{\mathrm{out}}}$ and the Brillouin Stokes (a) and anti-Stokes (b) signals, $\ket{\psi_{\mathrm{B}}}_{\mathrm{S/AS}}$.  
The value of the angle is indicated by the colorscale and is plotted as a function of the input polarization state $\ket{\psi_{\mathrm{in}}}$. 
} 
\label{figure3}
\end{figure}

Now, we study the influence of the input polarization state $\ket{\psi_{\mathrm{in}}}$ on the filtering of the laser while the micropillar characteristics ($\lambda_{\mathrm{cav},H/V}$, $\eta_{\mathrm{top},H/V}$ and $\kappa_{H/V}$) remain the same. 
The whole Poincaré sphere is covered by varying $\ket{\psi_{\mathrm{in}}}$.
For each given input polarization state, the maximal angle $\theta_{\mathrm{max}}$ reached between the two polarization states $\ket{\psi_{\mathrm{out}}}$ and $\ket{\psi_{\mathrm{B}}}_{\mathrm{S/AS}}$ in the Poincaré sphere is calculated. 
In Fig. \ref{figure3}, this maximal angle is plotted on the Poincaré sphere for the Stokes (Fig. \ref{figure3}.a) and the anti-Stokes (Fig. \ref{figure3}.b) signals. 
The coordinates of each point on the Poincaré sphere correspond to the Stokes parameters of the input polarization state $\ket{\psi_{\mathrm{in}}}$. 
The value of the maximal angle $\theta_{\mathrm{max}}$ is then given by the colorscale. 
It is important to note that $\theta_{\mathrm{max}}$ depends both on the incident polarization $\ket{\psi_{\mathrm{in}}}$ and on the laser wavelength $\lambda_{\mathrm{laser}}$. 
Therefore, in Fig. \ref{figure3}, the plotted value of $\theta_{\mathrm{max}}$ is  calculated for the laser wavelength which gives the best filtering, $\lambda_{\mathrm{opt}}^{\mathrm{S}}$ and $\lambda_{\mathrm{opt}}^{\mathrm{AS}}$, respectively. 
Hence, for each point, i.e., each $\ket{\psi_{\mathrm{in}}}$, $\theta_{\mathrm{max}}$ is reached for a different value of the input laser wavelength.

The minimal angle $\theta =0$ implies that  $ \ket{\psi_{\mathrm{out}}} = \ket{\psi_{\mathrm{B}}}_{\mathrm{S/AS}} $. 
It is is reached when the input polarization state is aligned with the cavity eigenmodes $\ket{H}$ or $\ket{V}$, as there is no polarization rotation induced by the cavity in these cases.  
Therefore, if $\ket{\psi_{\mathrm{in}}} = \ket{H/V}$, $\ket{\psi_{\mathrm{in}}} = \ket{\psi_{\mathrm{out}}} = \ket{\psi_{\mathrm{B}}}_{\mathrm{S/AS}}$.

The maximal angle $\theta_{\mathrm{max}}$ reached between the reflected laser polarization state and the Brillouin signals is $105^\circ$.
It is obtained for incident polarizations $\ket{\psi_{\mathrm{in}}}$ with coordinates in the Poincaré sphere corresponding to an annular shape around the HV axis and parallel to the DRAL meridian (in pink in Fig. \ref{figure3}). 
This maximal angle does not correspond to two opposite points in the Poincaré sphere.
Thus, a cross-polarization scheme, with the reflected laser polarization as reference, would completely filter out the laser but attenuate as well the Brillouin signal. 
For this given set of parameters ($\lambda_{\mathrm{cav},H/V}$, $\eta_{\mathrm{top},H/V}$ and $\kappa_{H/V}$), there is no input polarization state $\ket{\psi_{\mathrm{in}}}$ for which the states $\ket{\psi_{\mathrm{out}}}$ and $\ket{\psi_{\mathrm{B}}}_{\mathrm{S/AS}}$ are opposite to each other in the Poincaré sphere. \\

The parameters studied so far ($\lambda_{\mathrm{laser}}$ and $\ket{\psi_{\mathrm{in}}}$) can experimentally be tuned for a given micropillar to reach the maximal angle between the two polarization states $\ket{\psi_{\mathrm{out}}}$ and $\ket{\psi_{\mathrm{B}}}_{\mathrm{S/AS}}$. 
However, the orthogonality between those two polarization states cannot be reached in this case. 
We can overcome this limitation by varying the birefringence of the micropillar cross-section. 
Indeed, the splitting $\Delta \lambda$ between the two cavity modes, which has a determinant role in the rotation of polarization, is directly controlled by the ellipticity of the micropillar \cite{Gerhardt2019}. 
Experimentally, the birefringence is monitored by etching arrays of micropillars of varying ellipticity \cite{rodriguez_brillouin_2023}.
In the following, we study the effect of the ellipticity on the output polarization states by modifying $\Delta \lambda$ in the numerical simulations.

\begin{figure}[h!]
\centering
\includegraphics[scale=0.3]{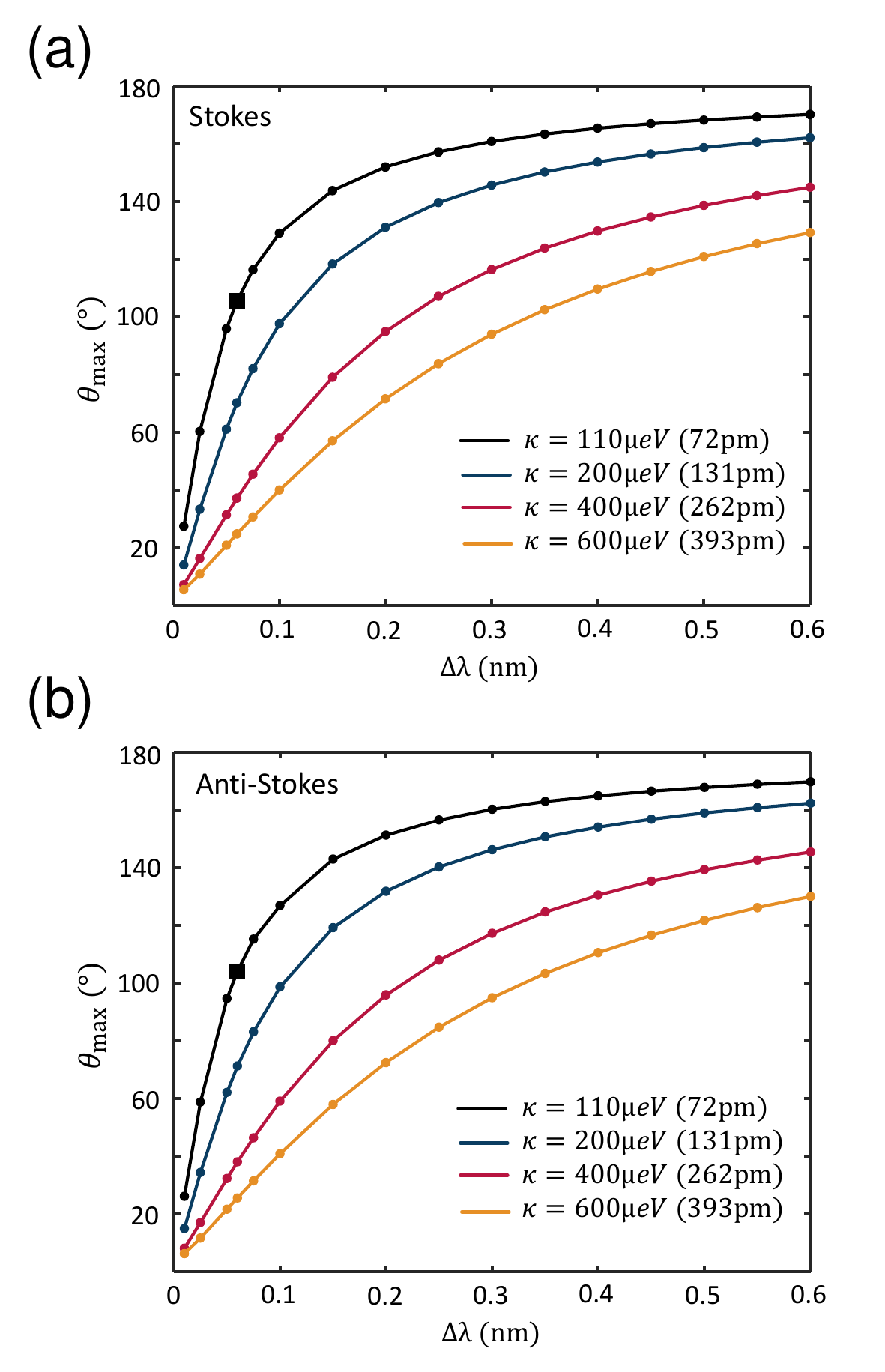}
\caption{
(a) and (b) Maximal angle $\theta_{\mathrm{max}}$ between $\ket{\psi_{\mathrm{out}}}$ and $\ket{\psi_{\mathrm{B}}}_{\mathrm{S/AS}}$ reached in the Poincaré sphere for a given $\ket{\psi_{\mathrm{in}}}$ for different values of $\kappa$, the total cavity damping rate.  }
\label{figure4}
\end{figure}

The maximal angle $\theta_{\mathrm{max}}$ in the Poincaré sphere, between the two polarization states $\ket{\psi_{\mathrm{out}}}$ and $\ket{\psi_{\mathrm{B}}}_{\mathrm{S/AS}}$, depends on how much the two optical cavity modes overlap in energy, which directly depends on $\Delta \lambda$, the separation between the central energies of the two optical cavity modes. 
In addition, the overlap between the two modes also depends on the full width at half maximum defined by the total cavity damping rate $\kappa_{H/V}$. 
A large overlap implies a small cavity-induced polarization rotation, and thus a small $\theta$. 
In contrast, a small mode overlap results in a more important rotation of polarization.
In Fig. \ref{figure4}.a and b, we plot the maximal angle $\theta_{\mathrm{max}}$ reached, associated to a given $\ket{\psi_{\mathrm{in}}}$ and $\lambda_{\mathrm{laser}}$, as a function of the cavity splitting between the two cavity modes $\Delta \lambda$ and for different values of the total cavity damping rate $\kappa$, where $\kappa_H = \kappa_V = \kappa$. 
The case where $\kappa = 110 \, \mu$eV (72$\,$pm) and $\Delta \lambda = 0.06\,$nm corresponds to the situation studied above and is indicated by a black square in Fig. 4. 
For each set of $\Delta \lambda$ and $\kappa$, the maximal value $\theta_{\mathrm{max}}$ is calculated. 
We observe that $\theta_{\mathrm{max}}$ increases with $\Delta \lambda$ and inversely with $\kappa$.
It reaches a maximum of $\theta_{\mathrm{max}} = 170^\circ$ for $\kappa = 110 \, \mu$eV (72$\,$pm). 
For low values of cavity damping rate $\kappa$, the spectral separation of the two modes increases rapidly with $\Delta \lambda$. Consequently, we will reach $\theta_{\mathrm{max}}$ and optimal conditions for a smaller ellipticity than in the case of a cavity with high damping rate. \\

Experimentally, an optimal configuration can be achieved for a given micropillar by tuning $\omega_{\mathrm{in}}$ and $\ket{\Psi_{\mathrm{in}}}$. 
In addition, the ellipticity of the micropillar cross-section can be designed through the fabrication process to control $\Delta\lambda$ and thus maximize the angle $\theta_{\mathrm{max}}$. 
The total cavity damping rate $\kappa_{H/V}$, on which the optimization of the filtering process depends as well, can also be designed with the fabrication process. 
Indeed, the number of layers in the DBRs controls the value of $\kappa_{H/V}$, yet limited by the micropillar fabrication.
Moreover, the spatial mode matching between the incoming laser beam and a micropillar can be experimentally limited by the ellipticity of the micropillar, which degrades the coupling between the input light and the cavity.
In such a situation, the efficiency and the intensity of the Brillouin scattering is compromised. 
Thus, finding the balance between an efficient mode matching between the optical cavity and the input laser, and the ellipticity of the cross-section is necessary to implement this polarization-based Brillouin spectroscopy technique.

\section{Conclusion}
We have theoretically studied the polarization control of elliptical optophononic micropillar resonators to optimize the Brillouin scattering detection.
We demonstrated that we can maximize the angle between the Brillouin signal and the excitation laser polarization states by independently manipulating them.
The reflected laser filtering can therefore be implemented with a cross-polarization scheme, while attenuating the Brillouin scattered signal. 
However, the numerical simulations presented in this work are restricted as they only take into account the rotation of polarization and not the intensity of the Brillouin scattering.
It is necessary to consider the enhancement of the Brillouin signal by the optical cavity.
This study can be experimentally applied to micropillars with different ellipticities, and be performed using a similar setup as presented in \cite{rodriguez_brillouin_2023}, adding the control of the incident polarization. 

This Brillouin spectroscopy technique can be extended to other polarization-dependent structures, such as nanofibers \cite{Zerbib2023}. 
Therefore, our findings are building blocks for efficient Brillouin detection in tunable optophononic resonators for future opto-mechanical applications. 

\section*{Acknowledgements}
The authors acknowledge funding from European Research Council Consolidator Grant No.101045089 (T-Recs). This work was supported by the European Commission in the form of the H2020 FET Proactive project No. 824140 (TOCHA), and the French RENATECH network.

%\bibliography{references.bib}

\end{document}